\documentclass[draft]{agujournal2019}
\usepackage{url} 
\usepackage{lineno}
\usepackage[inline]{trackchanges} 
\usepackage{soul}
\usepackage{amsmath}
\usepackage{amssymb}
\draftfalse

\journalname{Geophysical Research Letters}

\newcommand{\nat}{    {\it Nature}}
\newcommand{\apj}{    {\it The Astrophysical Journal}}
\newcommand{\jgra}{    {\it Journal of Geophysical Research: Space Physics}}

\newcommand{\grl}{    {\it Geophysical Research Letters}}
\newcommand{\solphys}{    {\it Solar Physics}}
\newcommand{\apjl}{    {\it The Astrophysical Journal Letters}}
\begin{document}

%
%


\title{On the quasi-three dimensional configuration of magnetic clouds}

%
%




\authors{Q. Hu\affil{1}, W. He\affil{2}, J. Qiu\affil{3}, A. Vourlidas\affil{4,5}, and C. Zhu\affil{3}}

\affiliation{1}{Department of Space Science, and
Center for Space Plasma and Aeronomic Research (CSPAR),
The University of Alabama in Huntsville,
Huntsville, AL 35805, USA}
\affiliation{2}{Department of Space Science,
The University of Alabama in Huntsville,
Huntsville, AL 35805, USA}
\affiliation{3}{Physics Department,
Montana State University,
Bozeman, MT 59717, USA}
\affiliation{4}{Johns Hopkins University Applied Physics Laboratory, Laurel, MD 20723, USA}
\affiliation{5}{IAASARS, National Observatory of Athens, GR-15236, Penteli, Greece}




\correspondingauthor{Qiang Hu}{qiang.hu@uah.edu}




\begin{keypoints}
\item First rigorous applications of a 3D model are carried out for in-situ measurements of MCs
\item Results via the optimal fitting approach yield reduced Chi2 values close to 1
\item Complexity of MC flux ropes is revealed by the model showing 3D winding magnetic flux bundles
\end{keypoints}

%
%

%
%


\begin{abstract}
We develop an optimization approach to model the magnetic field configuration of magnetic clouds, based on a linear-force free formulation in three dimensions. Such a solution, dubbed the Freidberg solution, is kin to the axi-symmetric Lundquist solution, but with more general ``helical symmetry''. The merit of our approach is demonstrated via its application to two case studies of in-situ measured magnetic clouds. Both yield results of reduced $\chi^2\approx 1$. Case 1 shows a winding flux rope configuration with one major polarity. Case 2 exhibits a  double-helix configuration with two flux bundles winding around each other and rooted on regions of mixed polarities. 
This study demonstrates the three-dimensional (3D) complexity of the magnetic cloud structures. 
\end{abstract}

\section*{Plain Language Summary}
Magnetic clouds (MCs) are a type of magnetic field structures observed in space. They possess some well-defined properties and have been well studied in the space age. The existing model for such a structure is a straight cylinder with no variation along its axis.  They may impact Earth carrying significant amount of electromagnetic energy.   They come in relatively large sizes. When encompassing the near-Earth space environment, their impact can last for days. MCs originate from the Sun, directly born with the so-called coronal mass ejections (CMEs) which can be seen as an ejection of large amount of solar material from telescopes aiming at the Sun. The CMEs are often accompanied by solar flares, the most energetic and explosive  events in our solar system. When these happen, they release a wide range of radiations and disturbances that may adversely impact Earth with MCs being one major type of such disturbances. Therefore studying the internal configuration of MCs is of importance to understanding their  origin and  impact. This study presents a more complex 3D MC model to better fit the in-situ spacecraft measurements of such structures, which goes beyond the current model.

%
%

\section{Motivation} \label{sec:intro}
Magnetic clouds (MCs) are  large-scale magnetic structures (usually with duration $\gtrsim$ 1 day at 1 au) observed from in-situ spacecraft measurements, such as those from the Advanced Composition Explorer (ACE) and Wind spacecraft in the solar wind. MCs possess three well-defined signatures in the magnetic field and plasma measurements: (1) relatively strong total magnetic field, (2) smooth rotation of one or more magnetic field components, and (3) depressed proton temperature or $\beta$ value (the ratio between the thermal  and  magnetic pressures). The elevated magnetic field and low $\beta$ value often indicate the dominance of the Lorentz force over the plasma pressure gradient and the inertia force for a magnetohydrostatic equilibrium. This leads to the force-free assumption such that the Lorentz force has to vanish. The simplest form, the linear force-free  field (LFFF) formulation, has been used to model the magnetic field configuration of  MCs. With one-dimensional (1D) dependence on the radial distance $r$ from a cylindrical axis only, the LFFF model yields the well-known Lundquist solution \cite{lund}, describing an axi-symmetric cylindrical flux rope configuration.   {MCs constitute a portion of interplanetary coronal mass ejections (ICMEs). A comprehensive study of the Wind spacecraft ICME Catalogue from 1995 to 2015 revealed the non-axisymmetric features of ICME flux ropes, and called for ``the development of more accurate in situ models''   \cite{2018SoPh..293...25N,2019SoPh..294...89N}. We intend to present such a model in this Letter, and will explore its wider applicability by applying to the Wind ICME Catalogue in a future study. }

{Besides a number of variations to the Lundquist solution
\cite<e.g.>[]{1999AIPCF,JGRA:JGRA52885}, which are mostly 1D (with $r$ dependence only),  \citeA{2016ApJ...823...27N} proposed a sophisticated circular-cylindrical model for MCs based on a generalized radial dependence of the current density. 
In addition,  the Grad-Shafranov (GS) reconstruction technique is able to obtain a 2D cross section of arbitrary shape  of a cylindrical structure based on single-spacecraft measurements \cite<see,>[for a comprehensive review]{Hu2017GSreview}. This method solves for the magnetic  flux function which defines distinct flux surfaces in a 2D configuration, governed by the GS equation.} The GS reconstruction was first applied to in-situ observations of magnetic flux ropes by \citeA{2001GeoRLHu,2002JGRAHu,Hu2003,2004JGRAHu}. The solution yields nested flux surfaces, representing winding magnetic field lines lying on distinct cylindrical surfaces surrounding a central straight field line.

MCs  are often entrained in coronal mass ejections (CMEs), and sometimes associated with solar flares. Efforts have been made to relate the MC flux rope configuration with the solar source region properties. {Specifically, we have carried out several investigations of comparing magnetic flux contents and field line twist profiles in MCs with those derived from flare observations, through in-situ modeling of MCs and the analysis of the magnetic reconnection sequences as manifested by the flare-ribbon brightenings in the source regions \cite{Qiu2007,2014ApJH,2017NatCo...8.1330W,2019ApJ...871...25W,zhu2020}.  These are largely based on highly quantitative observational analysis, with the understanding that magnetic reconnection (flare process) leads to the formation of the MC flux rope. Therefore the magnetic topology change during the flux rope formation process on the Sun, generally in three dimensions, contributes to the complexity of the internal structure of MCs. Numerous observations and numerical studies indicate the three-dimensional (3D) nature of flux rope configurations upon their origination on the Sun \cite<e.g.,>[]{2013SoPh..284..179V,2014PPCF...56f4001V,2018Natur.554..211A,2016NatCo...711522,2019ApJ...884...73D}, often in the form of twisted ribbons.} To account for such  features, we  develop an approach to probe the 3D MC field line configuration from in-situ data. An earlier attempt was made by \citeA{1999GeoRL..26..401O}, which showed a double-helix configuration as a solution to an alternative theoretical model, but lacked rigorous  applications to in-situ data. {That formulation takes a special form of a GS type equation, which we found to be difficult to apply to in-situ spacecraft measurements. Therefore the current approach reported here is developed to provide a new capability of modeling 3D MC structures by directly employing in-situ spacecraft measurements.}  


In what follows, we demonstrate our approach with optimal fitting of the 3D Freidberg solution \cite{freidberg} to single spacecraft measurements of MCs, strictly following the appropriate $\chi^2$ minimization methodology  \cite{2002nrca.book.....P}.  {In doing so, we intend to stimulate discussions on what defines a magnetic flux rope. As a general feature of the Freidberg solution as we reveal in the following sections, the magnetic field configuration deviates from a 2D geometry for a conventionally defined ``flux rope'' in that there generally does not exist a straight central field line. The field lines form flux bundles that wind along the $z$ dimension, similar to the topological feature of writhe as described in \citeA{1984JFMB}, and particularly by \citeA{2011ApJAl} for MCs. }

\section{Method} \label{sec:method} 
The method we develop is based on an LFFF formulation in three dimensions, namely, in a cylindrical coordinate system $(r,\theta,z)$. The following is a direct copy of the set of equations given in \citeA{freidberg}, representing a series solution to the equation $\nabla^2\mathbf{B}+\mu^2\mathbf{B}=0$ with the force-free constant $\mu$, 
\begin{eqnarray}
\frac{B_z(\mathbf{r})}{B_{z0}} & = & J_0(\mu r)+CJ_1(\alpha r)\cos(\theta+kz) \\
\frac{B_\theta(\mathbf{r})}{B_{z0}} & = & J_1(\mu r)-\frac{C}{\alpha}\left[\mu J'_1(\alpha r)+\frac{k}{\alpha r}J_1(\alpha r)\right]\cos(\theta+kz) \\
\frac{B_r(\mathbf{r})}{B_{z0}} & = & -\frac{C}{\alpha}\left[k J'_1(\alpha r)+\frac{\mu}{\alpha r}J_1(\alpha r)\right]\sin(\theta+kz). 
\end{eqnarray}\label{eq:B}
Such a solution (dubbed the Freidberg solution) is obtained by truncating the infinite series and keeping the first two modes through a standard separation of variables procedure. {For $C\equiv 0$, the solution reduces to the axis-symmetric Lundquist solution, and the traditional Lundquist solution fitting to MCs ensues.} Generally the solution has 3D dependence on spatial dimensions, but it is also periodic in $z$ with a period/wavelength $2\pi/k$, thus  called a solution of ``helical symmetry'' with mixed helical states of azimuthal wavenumbers $m=0$ and 1. The parameter $C$ determines the amplitude of the $m=1$ mode, which gives rise to the variation in $\theta$. Following \citeA{freidberg}, the LFFF constant is denoted $\mu$ and the parameter $\alpha=(\mu^2-k^2)^{1/2}$. The usual Bessel's functions of the first kind of the zeroth and  first order are denoted $J_0$ and $J_1$, respectively. {{The Freidberg solution has 3D variations in that the cross section varies along the $z$ dimension, which generally prohibits the appearance of a straight field line along $z$. Therefore for a ``flux rope'' configuration represented by the Freidberg solution, the writhe will be present in the form of winding flux bundles in lack of a central straight field line. }}

For an MC event detected in in-situ spacecraft data, an interval is chosen for a $\chi^2$ minimization process to determine the unknown parameters in  the Freidberg solution, i.e., equations~(1)-(3). A reduced $\chi^2$ function is defined to assess the difference between the measured magnetic field components $\mathbf{b}$ and the  analytic solution $\mathbf{B}$, subject to underlying uncertainties:
\begin{equation}
\chi^2=\frac{1}{\tt{dof}}\sum_{\nu=X,Y,Z}\sum_{i=1}^N\frac{(b_{\nu i} - B_{\nu
i})^2}{\sigma_{ i}^2}. 
\end{equation}\label{eq:chi2}
A minimum $\chi^2$ value is sought for an interval with $N$  magnetic field data points, often downsampled from 1-min cadence  to 1 hour. Then the degree of freedom ($\tt{dof}$) of the system is $3N-p-1$, with $p$ the number of parameters to be optimized. 
According to \citeA{2002nrca.book.....P}, a quantity $Q$,
indicating the probability of  a value greater than the
specific $\chi^2$ value, is also obtained for reference. It is calculated by $Q= 1 - \tt{chi2cdf}(\chi^2, \tt{dof})$, where the function $\tt{chi2cdf}$ is the cumulative distribution
function of $\chi^2$. The corresponding uncertainties $\sigma$ are estimated by taking the root-mean-square (RMS) variation of the underlying 1-min measurements over each one-hour interval, an approach adopted by the ACE Science Center MAG data processing (see \url{http://www.srl.caltech.edu/ACE/ASC/level2/mag_l2desc.html}). The set of main parameters to be optimized includes $C$, $\mu$, $k$, the pair of the directional angles of the $z$ axis, $(\delta,\phi)$, together with additional geometrical parameters to allow for more freedom of the solution with respect to the spacecraft path. Simply put, besides that the $z$ axis orientation is completely arbitrary, the outer cylinder enclosing the solution domain is allowed to translate along and perpendicular to, as well as to rotate about the  $z$ axis. This  fully accounts for the 3D nature of the solution. Detailed descriptions of the algorithm will be reported elsewhere. In the following case studies, the parameters $\mu$ and $k$ become dimensionless by multiplying a length scale $R_0$ which is the normalization constant for $r$ and $z$.

\begin{figure}
\centering
\includegraphics[width=8.3cm]{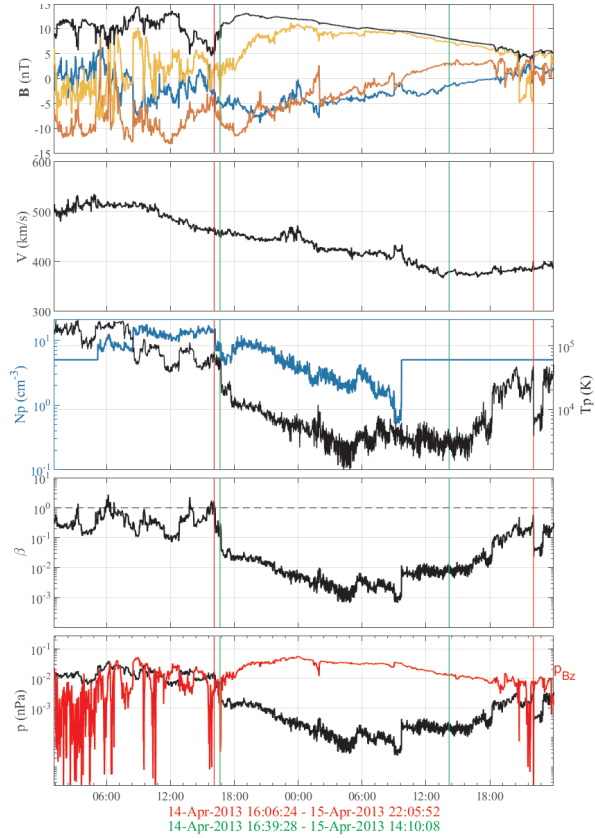}
\caption{Time-series from the ACE spacecraft measurements for Case 1. From  top to  bottom: magnetic field components in R (blue), T (brown), and N (gold) coordinates, and magnitude (black),  bulk speed,  proton density (left axis) and temperature (right axis),  proton $\beta$, and  thermal  and axial magnetic  pressure (red). The vertical lines mark the intervals for the GS reconstruction (green) and the optimization analysis (red) of the Freidberg solution  with the corresponding time periods denoted beneath the bottom panel, respectively.}\label{fig:dataplot}
\end{figure}

\section{Case Studies} \label{sec:results}
\begin{figure}
\centering
\includegraphics[width=8.3cm]{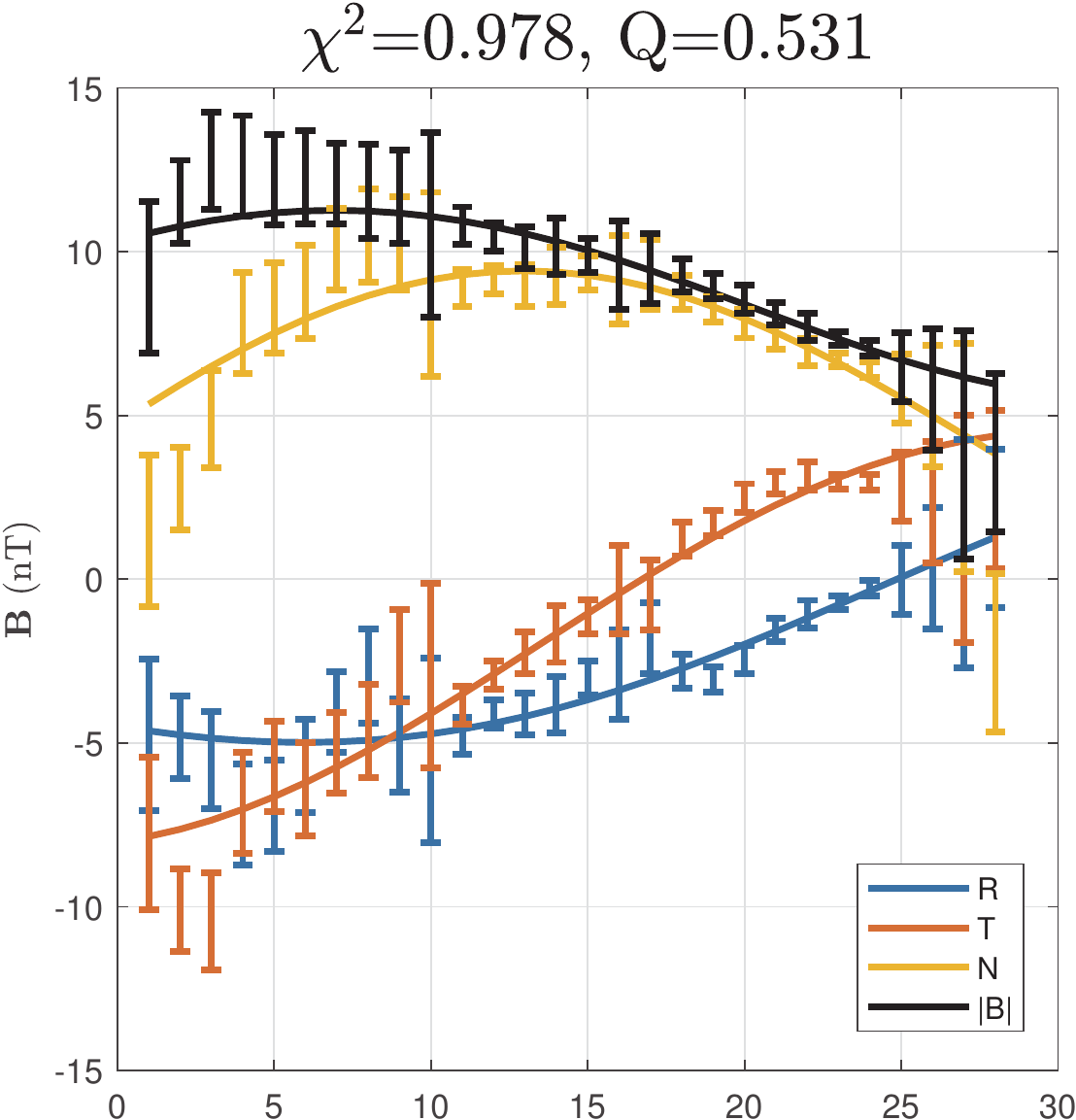}
\caption{The optimal fitting results to the Freidberg solution for Case 1. The error bars are the ACE measurements with uncertainties in hourly averages, and the solid curves are the Freidberg solution, for the R, T, N components, and the field magnitude, respectively, as indicated by the legend. }\label{fig:Brtn_r13}
\end{figure}

We present two case studies to illustrate the method. Case 1 is an MC event observed on 14-15 April 2013 at 1 au. Figure~\ref{fig:dataplot} shows the time-series plot from the ACE spacecraft measurements. A typical MC structure is present with relatively strong field magnitude and rotating field components, and depressed proton $\beta$. Two intervals are marked. Both last for over 20 hours. The average Alfv\'{e}n Mach number in the reference frame moving with the MC structure is 0.23, and the average $\beta$ is 0.01, justifying the assumption of quasi-static equilibrium and approximate force-freeness. 
A GS reconstruction was performed with  acceptable output. The optimization result for the Freidberg solution is shown in 
Figure~\ref{fig:Brtn_r13} with the minimum reduced $\chi^2=0.978$, and $Q=0.531$. 

{Table~\ref{tbl:para} lists the main fitting  parameters for the two cases. The normalization constants for the length scale and the magnetic field are denoted by $R_0$ and $B_{z0}$.  For the Freidberg solution, the parameter $C$ indicates the contribution from the variations in the $\theta$ and $z$ dimensions.   The parameter $k$ represents the wavenumber in the $z$ dimension. Therefore both the parameters $C$ and $k$ represent the 3D characteristics of the solution (for $C=0$, the solution returns to the 1D Lundquist solution, while for $k=0$, a 2D solution results). The force-free constant is given by $\mu$ and the sign of the parameter $\mu$ indicates the sign of magnetic  helicity (i.e., the handedness or chirality). The $z$ axis orientation is given by the polar and azimuthal angles $(\delta,\phi)$ in radians in the RTN coordinates. The axial magnetic flux within the positive polarity region (where $B_z>0$) on the cross section is denoted $\Phi_z$. }

{\centering
\begin{table}[h]

    \caption{Optimal fitting parameters of the Freidberg solution for the two case studies from the ACE spacecraft measurements.}\label{tbl:para}
    \begin{tabular}{cccccccc}
        \hline
         MC Interval (UT) &$R_0$ & $B_{z0}$  & $C$& $\mu$ & $k$ & $(\delta$, $\phi)$ & $\Phi_z$ \\
         hh:mm MM/DD/YY & AU & nT &			      &            &       &          Radians     &   $10^{20}$ Mx\\
\hline
       16:06 04/14/13 - 22:06 04/15/13 & 0.14  & 10.5 & 0.0367& -1.61 &  -1.60 & (0.433, 2.13) & 9.6\\
\hline
  08:04 07/15/12 - 13:52 07/16/12 & 0.33  & 21.9 &-2.27 & 5.64 & -4.07 & (0.867, 4.15) & 36\\
    \hline
    \end{tabular}
\end{table}}

%

\begin{figure}
\centering
\includegraphics[width=.45\textwidth]{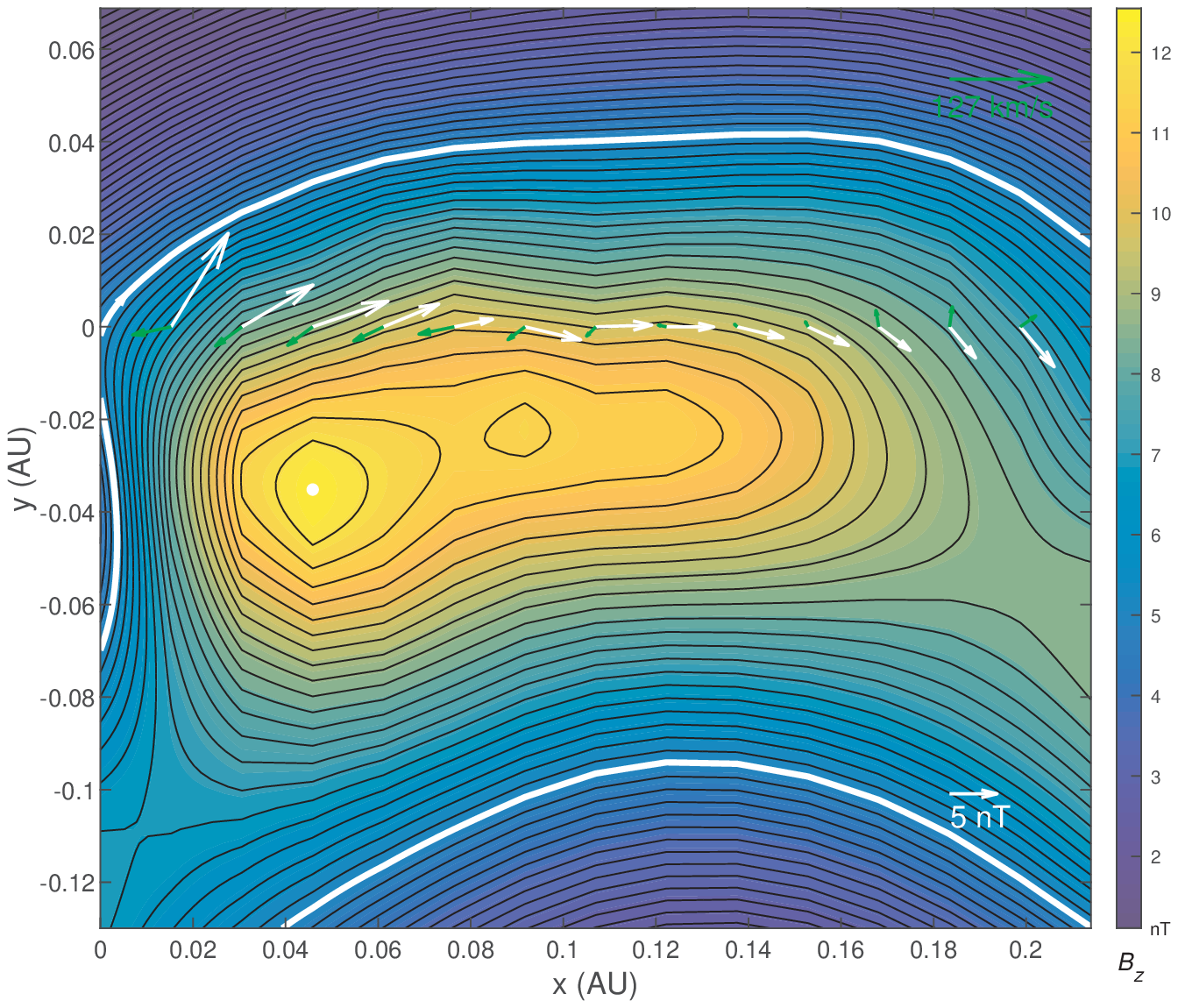}
\includegraphics[width=.53\textwidth]{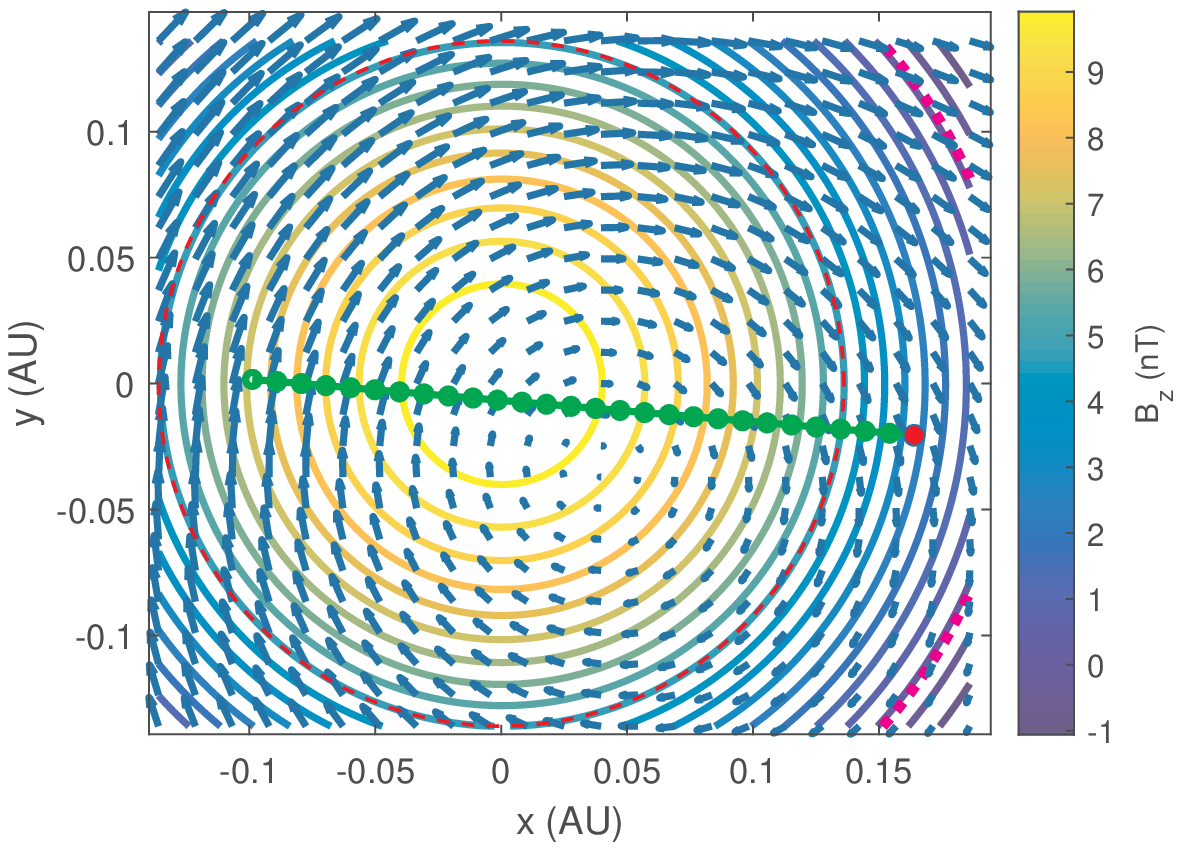}
\caption{The cross sections of the GS reconstruction result (left panel), and  the Freidberg solution at $z=0$ (right panel) for Case 1. In the left panel, the black contour lines represent the transverse field lines and color represents the axial field with scales indicated by the colorbar. The white (green) arrows along $y=0$ are the measured transverse field (remaining transverse flow) vectors along the spacecraft path. A reference vector for each set is shown (where the green reference vector is of the magnitude of the average Alfv\'en speed). In the right panel, the color contours show the axial field at $z=0$, and the corresponding transverse field is shown by arrows. The  dots mark the spacecraft path during the analysis interval in 1 hour increment from start (the leftmost green dot) to the end (the red dot). Note that they are not lying on this plane except for the leftmost dot. }\label{fig:maps}
\end{figure}

\begin{figure}
\centering
\includegraphics[width=.30\textwidth]{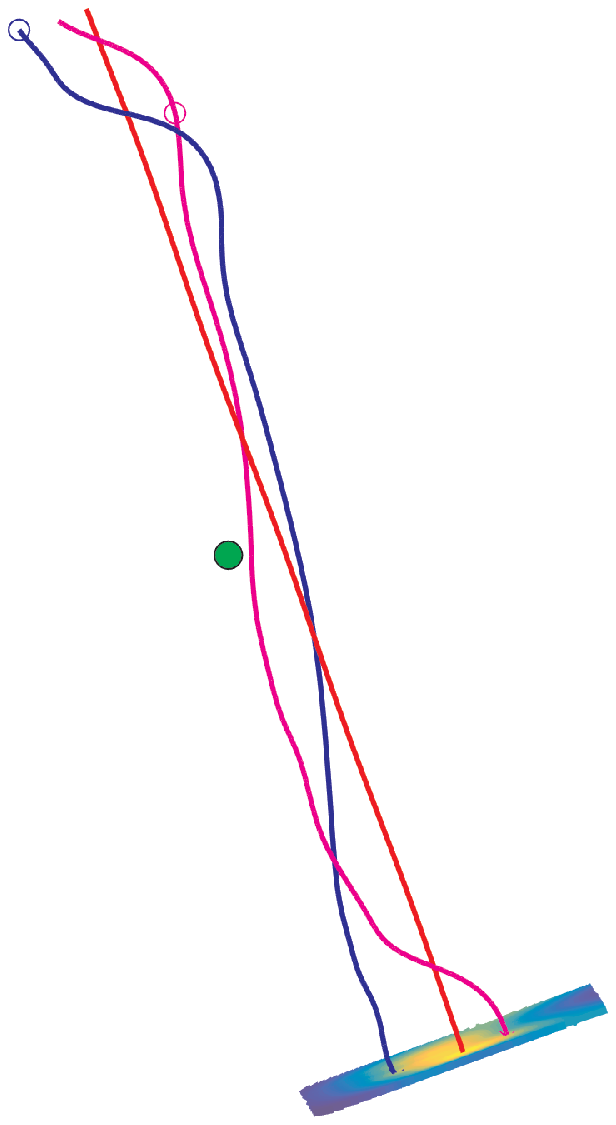}
\includegraphics[width=.18\textwidth]{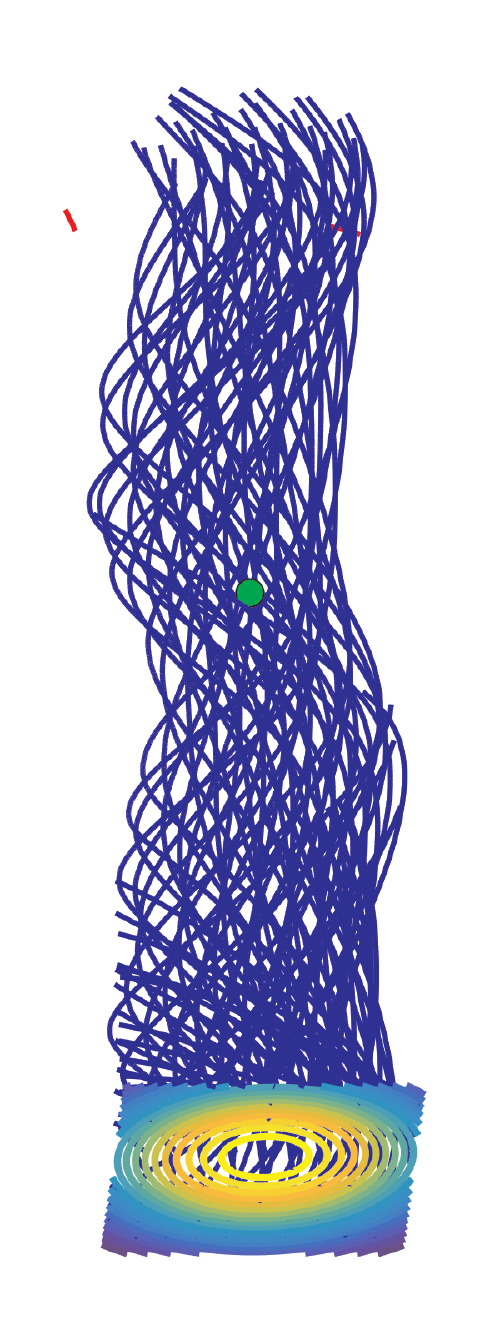}
\caption{The 3D view toward the Sun of the field line configurations for the GS reconstruction result (left panel), and the Freidberg solution (right panel), for Case 1. The big green dot marks the spacecraft path along the -R direction,  the N direction is straight up, and the T direction is horizontally to the right. Both sets of field lines are winding upward out of the bottom plane where contours of $B_z$ are shown. The $z$ axis orientations are (0.08206, -0.3377, 0.9377), and (-0.4706,-0.0350,0.8817), in RTN coordinates, respectively.}\label{fig:3D}
\end{figure}

Figures~\ref{fig:maps} and \ref{fig:3D} further demonstrate the similarity, but more pronounced the differences between the two solutions. Figure~\ref{fig:maps}, left panel, shows the cross section of a flux rope from the GS reconstruction in the form of the contour lines of the 2D flux function and the co-spatial axial field. In other words, the solution is fully represented by this 2D rendering in a view down the $z$ axis of a set of (nested) distinct flux surfaces. It is readily seen that the flux rope configuration is left-handed as indicated by the white arrows and the positive $B_z$ field along the spacecraft path. On the other hand, the Freidberg solution, given to the right, loses this 2D feature. This is the same view down the $z$ axis with the cross section drawn at $z=0$ where the first point along the spacecraft path is located. Then the spacecraft path (green dots) deviates from this plane. There are no distinct flux surfaces, and such a cross-section plot will change with $z$. Both solutions yield a uni-polar region of positive axial field and are left-handed. 
The axial magnetic flux is $\Phi_z$=$5.7\times10^{20}$ Mx, and $9.6\times10^{20}$ Mx, respectively. For the Freidberg solution, the sign of the parameter $\mu=-1.61$ indicates the negative sign of magnetic  helicity, i.e., left-handed chirality. The larger amount of flux in the Freidberg solution is partially  due to the corresponding larger interval used for this analysis (see Figure~\ref{fig:dataplot}). 

Figure~\ref{fig:3D} provides a 3D view of field line configurations toward the Sun for both solutions. Overall they are similarly oriented in space, with the $z$ axes pointing mainly northward. {The drastic difference, however, lies not in the number of field lines drawn for each, but in the intrinsic differences between a 2D and a (quasi-) 3D configuration}. In the right panel, more field lines are drawn to illustrate the overall winding of the flux rope body, which is not present in the left panel where the flux rope with a discernable central field line remains straight. 

%
  
\begin{figure}
\centering
\includegraphics[width=.45\textwidth]{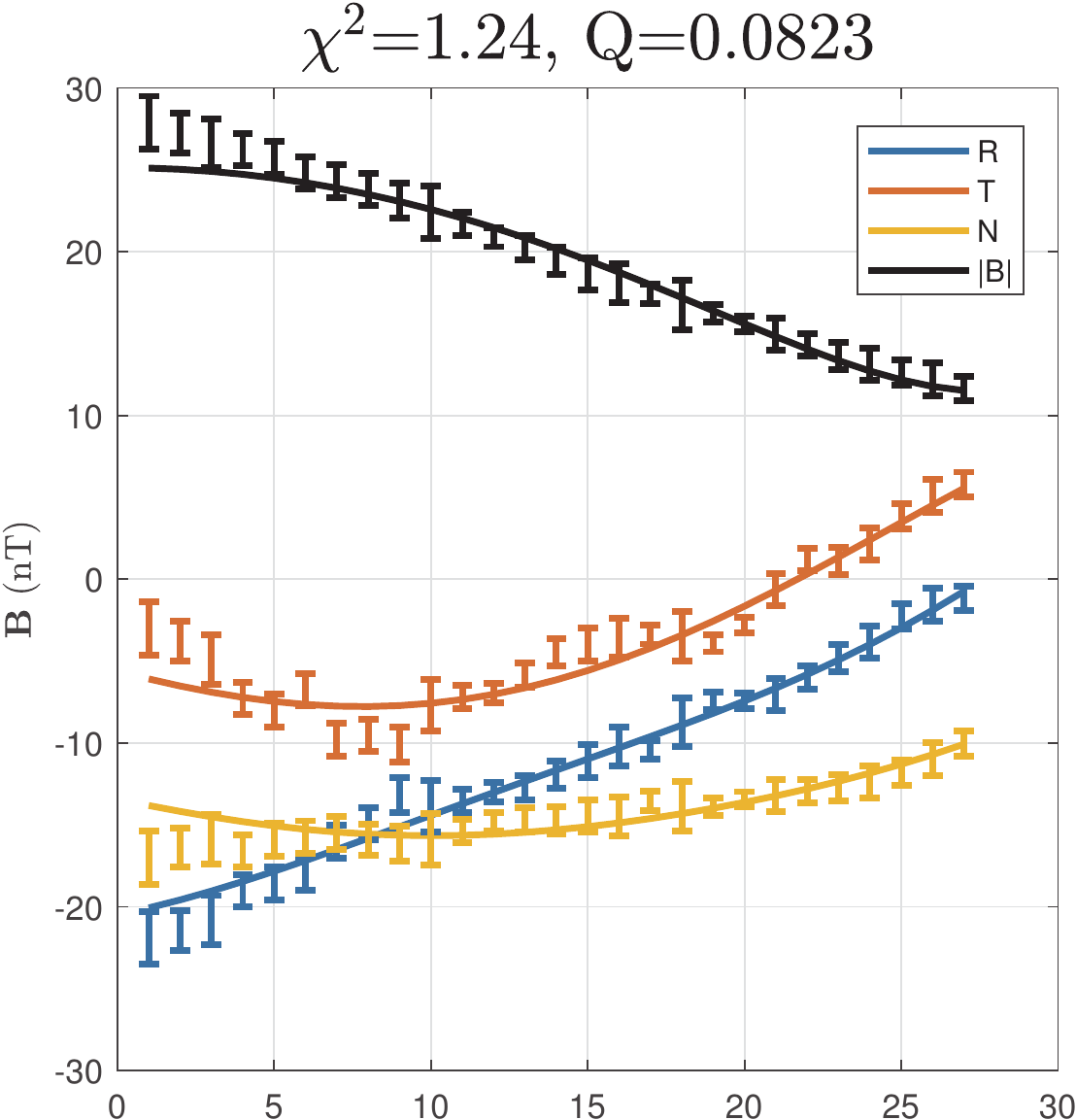}
\includegraphics[width=.54\textwidth]{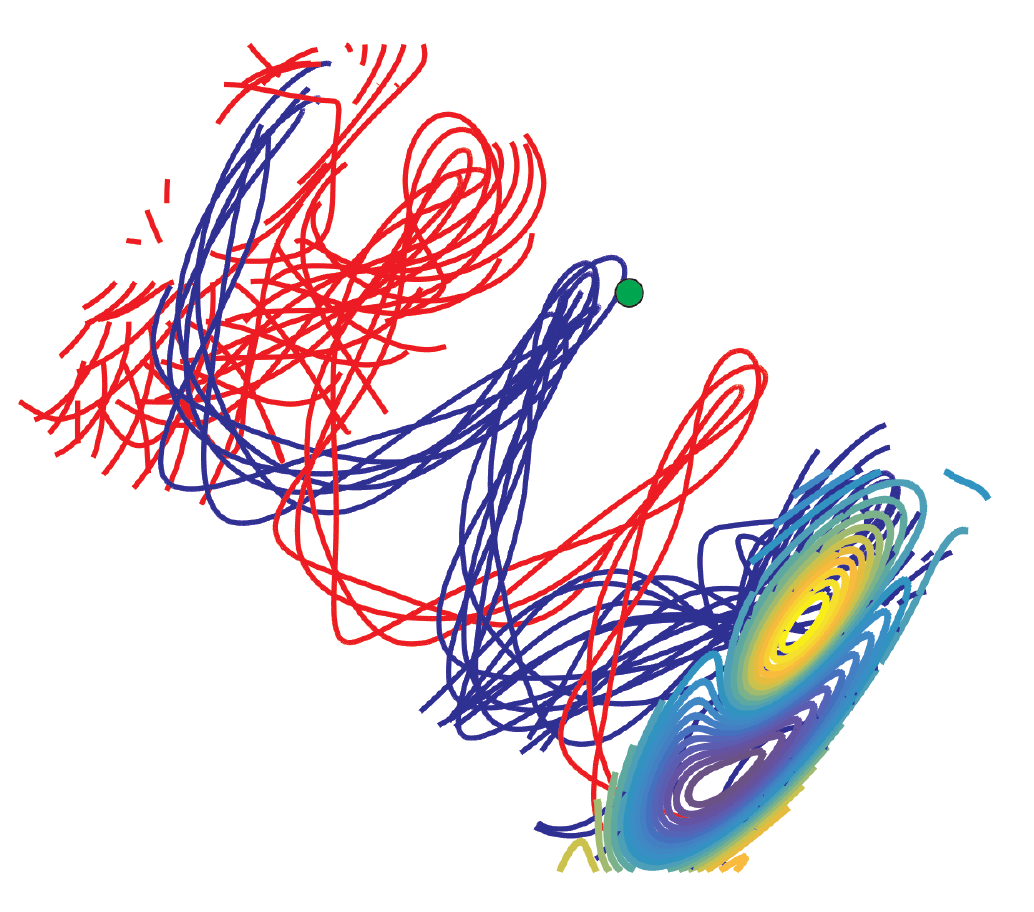}
\caption{Left panel: The optimal fitting result to the Freidberg solution for Case 2. The format is the same as Figure~\ref{fig:Brtn_r13}. Right panel: The 3D view toward the Sun of the field-line configuration for the Freidberg solution. The format is the same as Figure~\ref{fig:3D}. The set of red field lines are winding downward into the bottom plane. The $z$ axis orientation is $(-0.3265,-0.7509,0.5741)$ in the RTN coordinates. }\label{fig:case2}
\end{figure}

It is more informative to demonstrate by Case 2 the novelty of the new approach and the complexity of the field configuration represented by the Freidberg solution, whereas the GS reconstruction failed, {mainly due to the failure in finding a reliable invariance direction $\hat{\mathbf{z}}$ for a 2D configuration}. Case 2 is a well-studied Sun-Earth connection event with a prolonged MC interval  occurring on 15-16 July 2012. We refer readers to the VarSITI Campaign event webpage (\url{http://solar.gmu.edu/heliophysics/index.php/07/14/2012_17:00:00_UTC}) for detailed information and references on relevant studies.  An optimal Freidberg solution is obtained over a 27-hour  interval, as shown in the left panel of Figure~\ref{fig:case2}. The reduced $\chi^2$ value is slightly greater than 1. 
{The corresponding set of optimal parameters is given in the third row of Table~\ref{tbl:para}, indicating a more significant helical component ($|C|\gg 0$) and right-handed chirality ($\mu>0$). }
Indeed, the corresponding 3D field line configuration in Figure~\ref{fig:case2} (right panel) shows a striking double-helix structure with two bundles of field lines (blue and red) winding up and down along the $z$ axis and around each other. The cross section at the bottom clearly shows the mixed $B_z$ polarity regions next to each other, corresponding to the two flux bundles. Both are right-handed. In this event, the spacecraft is taking a glancing path across such a complex system.

\section{Summary} \label{sec:discussion}
In summary, we have developed a new approach to model the MC magnetic field in a quasi-3D configuration. The model is based on an LFFF formulation presented in \citeA{freidberg}, which is a generalization of the well-known Lundquist solution. The solution is 3D in nature as a function of $(r,\theta,z)$ in a cylindrical coordinate system, but with periodicity in $z$. A $\chi^2$ minimization process is devised by using the in-situ spacecraft measurements with underlying uncertainty estimates to determine the optimal set of parameters that yields a solution with the best fit to the magnetic field vectors along the spacecraft path. Two case studies are presented to illustrate the merit of the methodology. Both results are obtained with minimum reduced $\chi^2\approx 1$ and the associated $Q\gg 10^{-3}$, deemed acceptable according to \citeA{2002nrca.book.....P}. Case 1 exhibits a  flux rope configuration with  certain similarity to the corresponding 2D GS reconstruction result. Their $z$ axis orientations and  the axial magnetic flux contents are similar, and the chirality is the same. However the results are markedly different in that the Fieidberg solution exhibits a more general and intrinsicly 3D field configuration with a winding flux rope body. Potentially more complex MC structure is revealed by Case 2 in which a double-helix configuration is obtained. The cross section of the structure contains two adjacent regions of opposite field polarities (so are the currents) where the two helical flux bundles originate, both with right-handed chirality. Such a configuration, originating from the Sun, implies that the footpoint regions must have mixed polarities as well.  The ultimate proof of these  implications has to come from quantitative comparisons with solar source region properties. This future investigation involving more extensive lists of events with well-coordinated observations will be facilitated by this  new tool developed here, {complementary to the existing ones}, and will be pursued within our team.

\acknowledgments
The authors  acknowledge NASA grant 80NSSC18K0622 for partial support. We acknowledge useful discussions with Dr.~P.~Liewer. In addition,  QH acknowledges NASA grants 
80NSSC19K0276, 80NSSC17K0016 and NSF grant AGS-1954503 for  support. WH and QH acknowledge NSF grant
AGS-1650854 and NSO DKIST Ambassador program for support.  The ACE spacecraft merged magnetic field (MAG) and the solar wind electron, proton, and alpha monitor (SWEPAM) Level 2 data are publicly available via the ACE Science Center (\url{http://www.srl.caltech.edu/ACE/ASC/level2/lvl2DATA_MAG-SWEPAM.html}).


%
%


 \end{document}